# DESIGN AND PERFORMANCE OF A LOW-INTENSITY LED DRIVER FOR DETECTOR STUDY PURPOSES

G. Georgiev[1*], V. Kozhuharov[2], L. Tsankov

Faculty of Physics, University of Sofia "St. Kl. Ohridski", Sofia, Bulgaria

**Abstract.** *A custom LED driver producing light pulses with very low intensity and O(10 ns) duration was designed and constructed. A microcontroller was employed to handle the amplitudes and the repetition rates of the output pulses. In addition, it also provided both a PC control of the system through a RS232 interface and an external trigger I/O. A WLS fibre directly coupled to a LED provides unique characteristics of the output light pulse. The combination of a quasi delta light pulse source and physical absorption – emission medium results in an output light profile maximally close to the plastic scintillators. The light generator is intended to be used to test the response and the rate capability of different photodetectors. Its design, operational characteristics, and stability are described and discussed.*

*Key words*: *LED generator, short light pulses, scintillation detectors, photo detectors*

**DOI**: 10.21175/RadProc.2016.21

## 1. INTRODUCTION

During the design and the construction of any detector in field of particle physics it is important that the necessary equipment for test and calibration purposes is developed in parallel. Usually the adopted solution is rather custom made instead of commercial because of the large number of required channels. They are also specific for each particular case.

The scintillating detectors are commonly used to register ionizing radiation in scientific experiments. They consist of a light emitting material and a suitable photodetector. Generally, to test and calibrate the photodetector additional light sources are employed. Depending on the final goal various LED and laser light pulse generators are used. In most cases the pulse generator determines the output pulse width – in case of using a LED it could be constant of the order of 10 ns [1] (50 ns in the case of [2]) or with constant decay time (230 ns, as described in [3]). To obtain variable light pulse width at a nanosecond level usually more complicated solutions have to be developed [4]. Another possibility is to use laser source that is able to provide light pulses with a width of 70 ps [2]. However such a solution is more suitable for precise time resolution measurement than for response calibration since it provides a small dynamic range. The cost per single channel amounts to O(100) euros for custom made solution to O(1000) euros for commercial and not all the light pulse characteristics are customizable in general.

In the present work we describe a PC configurable light driver intended to be used for testing the response of the photodetectors (different type of PMTs, SiPM, CCD matrix) to the light emitted from plastic scintillators. Such a device is invaluable to address the detection capabilities of minimally ionizing particles at places where test beams with high energy particles are not available and when the active detector area and the corresponding cosmic rate is small. Moreover, the presented device is relatively inexpensive and is based only on commercially available components.

## 2. LED DRIVER DESIGN DESCRIPTION

The core operations of the LED driver are performed by the Microchip's 8-bit micro controller *PIC16F88*. It is a cheap but powerful device driven by main CPU with 386B Ram and 7kB programme memory. It is also equipped with ADC, DAC, comparators and communication modules – SPI, I²C and USART [5]. The main part of the code was written in *C* programming language but the time sensitive parts were written in *Assembler*. This way it was guaranteed that the external signals are with exact timing with respect to the light emission. A principle diagram of the device is shown in Fig. 1.

---

[1] Also at Theta Consult Ltd., Sofia, Bulgaria
[*] georgi.stefanov.georgiev@cern.ch
[2] Also at INFN – LNF, Frascati (Roma), Italy

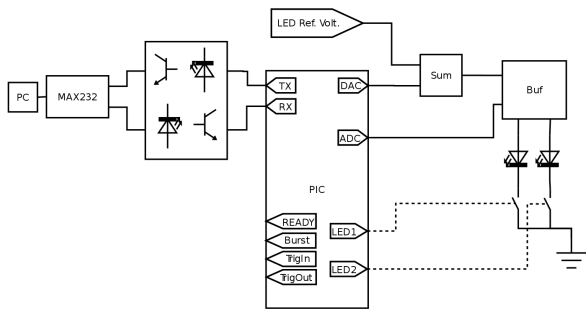

Figure 1. Diagram of the LED driver

*2.1. Communication module*

The communication is governed by the widely used protocol RS-232. The RS232 standard defines broad region for the voltage levels – from +3 V to +15 V and from -3 V to -15 V. MAX232 circuit is used for converting the RS232 voltage levels to MCU voltage levels (5 V) and vice versa. The built-in RS-232 module was utilized on the microcontroller's site. Two optron switches were placed between the microcontroller and the PC in order to reduce the induced noise in the communication line.

*2.2. Options and interface*

The generator is capable of generating sequences of short in time light pulses with predefined parameters. The number of pulses per sequence could vary from 0 to some tenths of thousands. The light intensity has 30 different programmable levels.

*2.3. IO signals*

The generator is intended to work autonomously. However, logical signals could trigger the sequence or could be transmitted to trigger an external device such as photo detector. Here is a full list of signals:

- READY (output) – "1" if the device is ready for next sequence, "0" otherwise;
- BURST (output) – "1" between the first and the last light pulse, "0" otherwise;
- TriggerOut (output) – Generates a pulse at the beginning of the sequence;
- TriggerIn (intput) – Triggers the sequence.

A jumper could be placed between TriggerIn and READY in order to work in continuous mode.

Two additional circuits were developed. The first one is basically a common collector which could be connected to one of the output TTL signals. This circuit gives possibility to use the TTL generated signal in 50Ω terminated external modules. The second one is driven by any of the TTL signals and produces negative pulses suitable for the discriminators in the common trigger systems.

*2.4. Electronic switch*

The electronic switch, schematically shown on fig. 1, was realized by the use of a fast n-p-n transistor and few additional components (fig. 2). The transistor base was connected to the TTL output of the microcontroller which generated a square pulse with duration of 1 µs. The emitter was grounded. The LED was connected between the collector and a capacitor. The circuit was powered by a stabilised voltage generator.

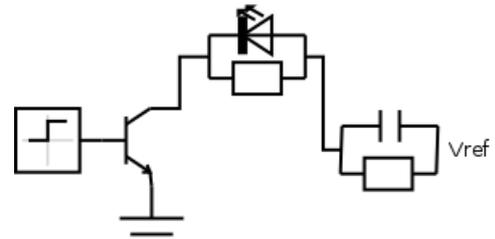

Figure 2. Electronic switch schematics

Once a positive voltage (e.g. TTL "1") is applied to the base, the charge passing through the LED is $Q=CU_{ref}$.

Two additional resistors are used in parallel to the capacitor and the LED. The RC circuit constant is some µs and is meant to discharge the capacitor out of the pulse. The LED and the resistor in parallel could be connected to the opposite side of a relatively long symmetric transfer line. The resistor reduces the impact of the noise and ensures a higher current in the line.

*2.5. Feedback and intensity control*

The intensity of the emitted light is controllable by change of the reference voltage set to the electronic switch. The voltage stabilizer was built using two operational amplifiers *TL082* and the variable resistor matrix incorporated in the microcontroller, exploited for digital to analog signal conversion.

An additional LED of the same kind connected in a static regime sets the bias reference voltage for the pulser LEDs. The bias was summed with the voltage generated by the DAC in one of the operational amplifiers. The ratio of the summator inputs was tuned. The second operational amplifier in the package was used as a buffer.

The ADC incorporated in the microcontroller was used to digitize the buffer output voltage for feedback purposes.

The early versions of the firmware were written completely in C language. This resulted in degraded performance, causing a delay of several micro seconds between the output TTL signals (Trigger, Burst) and the actual emission of the light pulse. Substituting the corresponding part of the C code with a low level programming and setting the level of the TTL outputs one instruction in advance decreased the delay corresponding to one instruction clock (1 µs, for 4 MHz main clock). The jitter of the electronic switch signal with respect to the output TTL signals was improved from 30 ns (using only the PIC*16F88* internal clock generator) to 1 ns by implementing a quartz resonator for frequency stabilization.

*2.5. Light transmission*

The short LED flashes were transferred to a WLS or scintillating fibre, glued to the LED lens. The scintillating fibre acts as a waveguide while the WLS fibre absorbs and re-emits the blue light towards the green part of the spectrum. When the time width of the LED light pulse is shorter with respect to the WLS fibre decay constant the light intensity profile at the output is dominated by the WLS characteristics.

## 3. SIGNAL CHARACTERISTICS

The main goal in the construction of the LED driver was to resemble the light pulse from a plastic scintillator. The characteristics of the designed equipment were studied with a specially developed experimental system, schematically shown in Fig. 3.

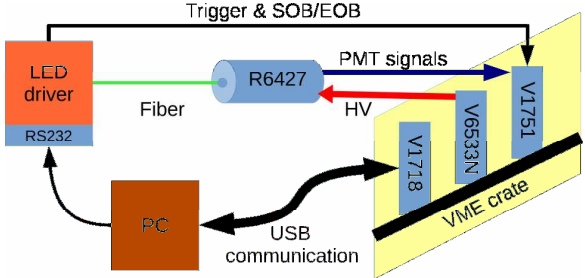

Figure 3. Schematics of the setup used to study the performance of the developed LED driver

It consisted of:

- LED driver, as described in section 2.
- Reference PMT: A 10-stage 25-mm Hamamatsu R6427 tube [6] was used as a reference photomultiplier. It features a 1.7 ns signal rise time. The PMT divider was E2624-04 and the gain was approximated as

$$G = 1.497 \times U^{6.784} \times 10^{-15}, \quad (1)$$

where U is the applied high voltage (in volts).

- HV module: CAEN V6533N [7] provided the negative voltage to the PMT divider. The voltage was controlled manually through the CAENGECO software.
- Digitizer: A CAEN V1751 digitizer [8] was used to convert the analog signal into digital data. It has 8 channels sampled continuously at 1 GS/s with 10 bit resolution. The dynamic range of V1751 is 1 V and special care was taken to prevent the inputs of the digitizer from damage.
- V1718 crate controller [9]. The crate controller communicated with a PC via its USB interface.
- PC. The control PC had two major functions. It provided the necessary configuration to the LED driver and it also served as a data acquisition PC running a custom developed DAQ program, collecting the data from the V1751 digitizer.

The data was stored in binary format containing all the samples in a given window with respect to the arrival of an external trigger. The trigger signal was provided by the LED driver itself. The signal from the PMT was assumed to be around the sample with the maximal (in absolute value) amplitude. An event-by-event pedestal was subtracted from each sample, estimated from several samples far from the PMT signal. The total charge and the number of the photoelectrons were calculated according to

$$N_{ph.e^-} = \frac{\int_{pulse} V(t)dt}{RGq}, \quad (2)$$

where V(t) are the samples of the amplitude from PMT, R is the impedance of digitiser, G – the gain of the PMT, q – elementary charge.

A number of blue LEDs coupled to wavelength shifting (WLS) fibres or scintillating fibres were used. The WLS fibers were 1mm BCF-92 with ~3ns decay time, and the scintilling ones were 1mm Kuraray SCSF-81 multiclad fibres. While the time profile of the light pulse at the SCSF-81 fibre end is mostly determined by the light emission of the photodiode and the light propagation along the fibre, the BCF-92 fibre initially absorbs the blue light by activating the dissolved fluor centres, then the excited molecules go to a lower energy state emitting green light with a given decay constant. That is why the time profile of the light at the end of the BCF-92 is almost completely dominated by the WLS characteristics.

During the tests data were collected with both the WLS and scintillating fibers, varying the voltage of the LED. An example of the signals from the PMT for different LED pulse voltages is shown in Fig. 4. The trigger signal with a maximal jitter <1ns defining the start of the pulse is also visible.

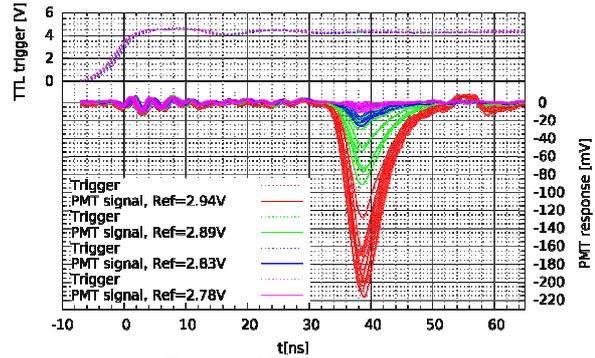

Figure 4. LED driver regimes

The parameters of the LED-driver – PMT assembly were compared with data from cosmic rays. Muons passed through a block of EJ-212 plastic scintillator with dimensions 3.8×3.8×10cm³ and the waveforms were recorded [10]. The signal reconstruction was done in the same way as for the LED driver. An example of the superimposed signals from EJ-212 exposed to cosmic rays and the LED-driver is shown in Fig. 5. The signals were normalized to maximal amplitude of 100 mV.

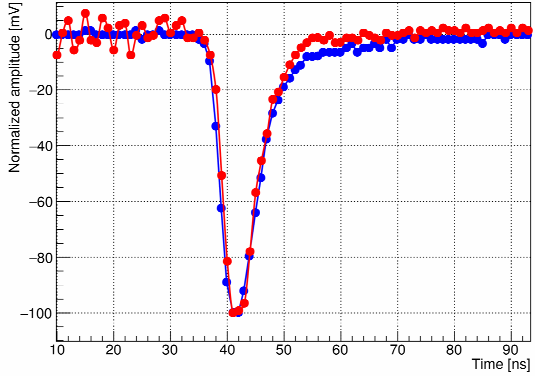

Figure 5. The signal shape from the R6427 PMT with the LED-driver (WLS fibre, in red) compared to the signal shape from EJ-212 scintillator (in blue)

The most important property of the reconstructed signal shape is the fall time of the PMT pulse. It was defined as the time necessary for the signal to decrease from 90% to 10% of its maximal amplitude. For PMTs with fast rise time (as is the case for R6427) and high bandwidth of the readout system, the fall time is mostly dependent on the characteristics of the light emitting medium. Pulse samples were collected with WLS and scintillating fibers, and compared with the fall time distribution for pulses caused by cosmic ray interactions in the EJ-212 scintillator. The results are shown in Fig. 6. The mean fall times for the different configurations were 6.5ns, 8.1ns, and 8.8ns for WLS, scintillating fibre, and cosmics on EJ-212, respectively. As expected, the pulse shape from the LED driver coupled with the WLS fibre resembles best the time properties of the signal, caused by the light emission of EJ-212 due to ionizing radiation. It should be noted that the average number of photoelectrons (and thus the amplitude of the signal) in the EJ-212 sample varies a lot because of the different possible paths of the cosmic rays through the active volume of the scintillator, leading to somewhat higher dispersion of the properties of the signals.

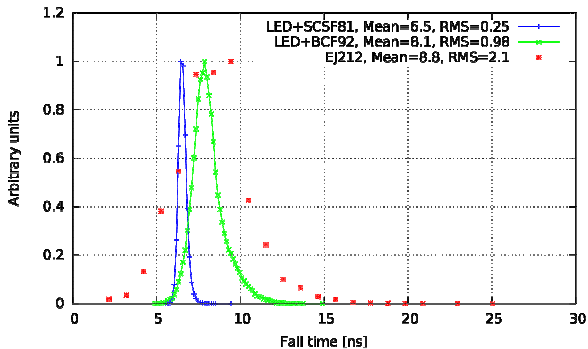

Figure 6. Distribution of fall times for WLS and scintillating fibers

## 4. STABILITY AND LIGHT YIELD

The stability of the operation of the LED driver was checked performing long runs with a 1000 pulses per a burst. The individual pulses were separated by 256μs. A start and end of burst signals wrapping the pulse train were used to start the DAQ and each burst was recorded in a separate file. The total time for a single burst including the initialization time both for the DAQ and the LED-driver and the postprocessing time was ~42s. More than 120 bursts were recorded for each of the configurations studied.

The chosen testing setup allowed checking both the time stability of the light emission, shown for the WLS fibre in Fig. 7, and the stability of the pulse amplitudes within the burst itself, shown in Fig. 8.

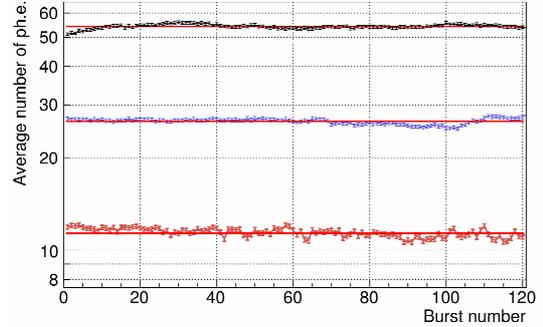

Figure 7. Stability of the average number of photoelectrons during a continuous operation, for three different LED switching voltages

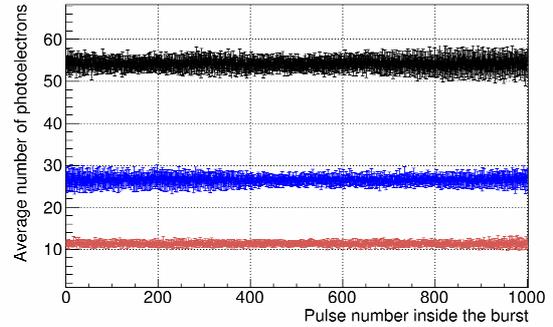

Figure 8. Stability of the average number of photoelectrons with the pulse number in the burst

A variation in the light yield of the order of 1% was observed, more expressed for low light intensity when the LED operates close to its threshold. The observed RMS of the variation of the average number of the photoelectrons $N_{pe}$ with time was attributed to a systematic uncertainty to the $<N_{pe}>$. There is no observable variation in the pulse amplitude between the first and the last pulse in a given burst and the change in $<N_{pe}>$ affects all the 1000 pulses in an equal way. This change with time was attributed to the lack of temperature stabilization of any part of the developed circuit and particularly of the reference voltage.

Table 1. Number of photoelectrons measured with WLS fibre and scintillating fibre

| $V_{ref}$ | $N_{pe}^{wls}$ | $R^{wls}$ | $N_{pe}^{sc}$ | $R^{sc}$ |
|---|---|---|---|---|
| 2.72 | | | 0.94±0.1 | 1.13 |
| 2.78 | | | 3.16±0.12 | 3.58 |
| 2.83 | | | 14.7±0.2 | 9.96 |
| 2.89 | 3.70±0.12 | 2.66 | 47.5±0.27 | 27.3 |
| 2.94 | 11.3±0.37 | 6.06 | 137±0.61 | 65.4 |
| 3.00 | 26.5±0.57 | 10.6 | 305±1.5 | 121 |
| 3.05 | 54.1±0.75 | 18.6 | | |

Since the variation of the average amplitude is small, a calibration of the LED driver was performed. The values are listed in Table 1 which shows the mean value <$N_{pe}$> and the RMS ($R^{wls/sc}$) of the number of the photoelectrons emitted from the photocathode of the PMT. The RMS describes the dispersion of <$N_{pe}$> for a single pulse. While the RMS is relatively high, the average is known with a precision of about 1 photoelectron.

The relation between the average number of the emitted photoelectrons both for the WLS fibre and the scintillating fibre is shown in Fig. 9. The exponential dependence of <$N_{pe}$> on $V_{ref}$ is clearly visible. For the same $V_{ref}$, the value of <$N_{pe}^{wls}$> is smaller than <$N_{pe}^{wls}$> due to the capturing efficiency of the WLS fibre, estimated to be ~8%.

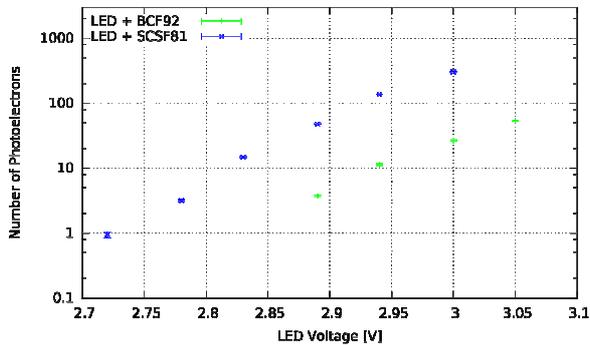

Figure 9. Number of photoelectrons as a function of LED voltage

5. CONCLUSIONS AND DISCUSSIONS

The proposed device described in the present paper is simple and capable to generate very short in time light pulses. The reproduction of the proper light profile to be as close as possible to the emission from the physical detector plays a crucial role when addressing the high rate response and double pulse separation capabilities. This was achieved by coupling a WLS fibre to the LED since the life time of the excited states of fluors is larger than the duration of the LED light pulse. The light generator provides stable with time pulses and was calibrated with a reference PMT up to O(100) photoelectrons. The temperature stability is of the order of 5 % and is expected to be improved with the introduction of a compensating circuit. Despite the lack of possibility to vary the signal profile in the case of the usage of WLS fibre the described light driver still has broad application due the wide spread of plastic scintillator detectors.

***Acknowledgments.*** *The present work was partially supported by the University of Sofia science fund under grants FNI-SU N57/2015 and FNI-SU N39/2016, and by Theta Consult Ltd. The authors acknowledge Dr. Ilko Rusinov for the useful discussions and suggestions, Dr. Roumen Tsenov for providing part of the used equipment, and LNF-INFN for the technical support under LNF-SU contract 70-06-497/07-10-2014. VK and GG also acknowledge Dr. Leandar Litov for the possibility to use part of the equipment at the HEP laboratory at the University of Sofia.*